\documentclass[apj]{emulateapj}

\newcommand{\bm}[1]{\mbox{\boldmath$#1$}}
\newcommand{\be}{\begin{equation}}
\newcommand{\ee}{\end{equation}}
\newcommand{\etal}{et al.}
\newcommand{\mpr}{m_\mathrm{p}}
\newcommand{\zg}{z_\mathrm{g}}
\newcommand{\rg}{r_\mathrm{g}}
\newcommand{\Teff}{T_\mathrm{eff}}

\newcommand{\Rinfty}{R^\infty}
\newcommand{\ThetaB}{\Theta_B}
\newcommand{\vecB}{\bm{B}}
\newcommand{\veck}{\bm{k}}
\newcommand{\veckp}{\bm{k'}}
\newcommand{\thetak}{\theta_k}
\newcommand{\phik}{\phi_k}
\newcommand{\vecn}{\bm{n}}
\newcommand{\vecm}{\hat{\bm{m}}}
\newcommand{\thetam}{\theta_m}

\newcommand{\Bpole}{B_{\mathrm{pole}}}
\newcommand{\xspec}{XSPEC}
\newcommand{\nsmax}{NSMAX}
\newcommand{\nsa}{NSA}
\newcommand{\nsagrav}{NSAGRAV}
\newcommand{\nsatmos}{NSATMOS}
\newcommand{\nsspec}{NSSPEC}
\newcommand{\fluxnorm}{A}

\shorttitle{Spectra of Neutron Stars with Magnetic Atmospheres}
\shortauthors{Ho, Potekhin, \& Chabrier}
 
\begin{document}

\title{Model X-ray Spectra of Magnetic Neutron Stars with Hydrogen Atmospheres}
                                                                                
\author{Wynn C. G. Ho}
\affil{Harvard-Smithsonian Center for Astrophysics,
60 Garden St., Cambridge, MA, 02138, USA
\\\texttt{wynnho@slac.stanford.edu}}
\and
\author{Alexander Y. Potekhin\altaffilmark{1}}
\affil{Ioffe Physico-Technical Institute,
Politekhnicheskaya 26, 194021 St.\ Petersburg, Russia
\\\texttt{palex@astro.ioffe.ru}}
\and
\author{Gilles Chabrier}
\affil{Ecole Normale Sup\'{e}rieure de Lyon, CRAL (UMR CNRS No.\ 5574), 
69364 Lyon Cedex 07, France
\\\texttt{chabrier@ens-lyon.fr}}
\altaffiltext{1}{Also at the Isaac Newton Institute of Chile 
(St.\ Petersburg Branch), St.\ Petersburg, Russia}

\begin{abstract}
We construct partially ionized hydrogen atmosphere models for magnetized
neutron stars in radiative equilibrium with fixed surface fields between
$B=10^{12}$ and $2\times 10^{13}$~G and effective temperatures
$\log\Teff=5.5$--$6.8$, as well as with surface $\vecB$ and $\Teff$
distributions around these values.
The models are based on the latest equation of state and opacity results
for magnetized, partially ionized hydrogen plasmas.
The atmospheres directly determine the characteristics of thermal
emission from the surface of neutron stars.
We also incorporate these model spectra into \xspec, under the model
name \nsmax, thus allowing them to be used by the community to fit
X-ray observations of neutron stars.
\end{abstract}


\section{Introduction} \label{sec:intro}

Thermal radiation has been detected from radio pulsars and radio-quiet
neutron stars (NSs; see
\citealt*{kaspietal06,haberl07,vankerkwijkkaplan07,zavlin07}, for reviews)
and from soft gamma-ray repeaters and anomalous X-ray pulsars, which form
the magnetar class of NSs endowed with superstrong ($B\gtrsim 10^{14}$~G)
magnetic fields (see \citealt{woodsthompson06}, for a review).
Radiation from the surface of these NSs can provide invaluable
information on the physical properties and evolution of the NSs.
Characteristics of the NS, such as the gravitational mass $M$,
circumferential radius $R$, and surface temperature $T$,
depend on the poorly constrained physics of the stellar interior,
such as the nuclear equation of state (EOS) and
quark and superfluid/superconducting properties at supra-nuclear densities.
Many NSs are also known to possess strong magnetic fields
($B \sim 10^{12}$--$10^{13}$~G), with some well above the quantum critical
value ($B\gg B_\mathrm{Q} = 4.414\times 10^{13}$~G).

The observed radiation from a NS originates in a thin atmospheric
layer (with scale height $\sim 1$~cm)
that covers the stellar surface.
To properly interpret the observations of NS surface emission and to
provide accurate constraints on the physical properties of NSs, it is
important to understand in detail the radiative behavior of NS
atmospheres in the presence of strong magnetic fields.
The properties of the atmosphere, such as the chemical composition, EOS,
and radiative opacities, directly determine the characteristics of the
observed spectrum.
While the surface composition of the NS is unknown, a great simplification
arises due to the efficient gravitational separation of light and heavy
elements (see \citealt*{alcockillarionov80,brownetal02}).
A pure hydrogen atmosphere is expected even if a small amount of
fallback/accretion occurs after NS formation;
the total mass of hydrogen needed to form an optically thick atmosphere
can be less than $\sim 10^{16}$~g.
Alternatively, a helium atmosphere may be possible as a result of
diffusive hydrogen burning on the NS surface
\citep*{changbildsten03,changetal04}.
Finally, a heavy element atmosphere may exist if no accretion takes place
or if all the accreted matter is consumed by thermonuclear reactions.
 
Steady progress has been made in modeling NS atmospheres
(see \citealt*{pavlovetal95,holai01,holai03,zavlin07}, for more detailed
discussion and references on NS atmosphere modeling).
Since the NS surface emission is thermal in nature, it has been modeled
at the lowest approximation with a blackbody spectrum.
Early works on atmospheric spectra assume emission from light element,
unmagnetized atmospheres (the latter assumption being valid for
$B\lesssim 10^9$~G); computed spectra exhibit significant deviation
from a Planckian shape and distinctive hardening with respect to a blackbody
\citep*{romani87,rajagopalromani96,zavlinetal96,gansickeetal02}.

The strong magnetic fields present in NS atmospheres significantly
increase the binding energies of atoms, molecules, and other bound states
(see, e.g., \citealt{lai01}, for a review).
Abundances of these bound states can be appreciable in the atmospheres of
cold NSs (i.e., those with surface temperature $T\lesssim 10^6$~K;
 \citealt*{laisalpeter97,potekhinetal99}).
In addition, the presence of a magnetic field causes emission to be
anisotropic \citep{pavlovetal94,zavlinetal95a}
and polarized \citep{meszarosetal88,pavlovzavlin00};
this must be taken into account self-consistently when developing
radiative transfer codes.
The most comprehensive early studies of magnetic NS
atmospheres focused on a fully ionized hydrogen plasma and
moderate field strengths ($B\sim 10^{12}$--$10^{13}$~G; 
\citealt*{miller92,shibanovetal92,pavlovetal94,zaneetal00}).
These models are expected to be valid only for relatively high temperatures
($T\gtrsim \textrm{a~few}\times 10^6$~K), where hydrogen is almost completely
ionized.
More recently, atmosphere models in the ultra-strong field
($B\gtrsim 10^{14}$~G) and relevant temperature regimes have been presented
(\citealt{holai01,holai03,ozel01,zaneetal01,lloyd03,vanadelsberglai06};
see \citealt{bezchastnovetal96,bulikmiller97},
for early work), and all of these rely on the assumption of a fully
ionized hydrogen composition (see, however, \citealt{hoetal03}).
Magnetized non-hydrogen atmospheres have been studied by \citet{miller92}
and \citet*{rajagopaletal97}, but because of the complexity of the
atomic physics, the models were necessarily crude (see \citealt{moriho07},
for more details).  Only recently have self-consistent atmosphere models 
\citep{hoetal03,potekhinetal04,moriho07} using the latest EOS and opacities
for partially ionized hydrogen \citep{potekhinchabrier03,potekhinchabrier04}
and mid-$Z$ elements \citep{morihailey02,morihailey06} been constructed.

Here we present a systematic tabulation of our
partially ionized hydrogen atmosphere models
for $B=10^{12}$--$2\times 10^{13}$~G.
We incorporate these tables into
\xspec\footnote{http://heasarc.gsfc.nasa.gov/docs/xanadu/xspec/}
\citep{arnaud96}, under the model name \nsmax,
for use by the astronomical community.
Note that the NS atmosphere models previously implemented in \xspec\
are either non-magnetic (\nsagrav: \citealt{zavlinetal96};
\nsspec: \citealt{gansickeetal02};
\nsatmos: \citealt*{mcclintocketal04,heinkeetal06})
or magnetic but fully ionized hydrogen 
(\nsa: \citealt{pavlovetal95});
the last at two fields: $B=10^{12}$~G and $10^{13}$~G.
In \S~\ref{sec:atmmodel}, we give details on the construction of the
atmosphere models.
In \S~\ref{sec:results}, we present our results, and we summarize
and mention future work in \S~\ref{sec:discussion}.

\section{Construction of Atmosphere Model} \label{sec:atmmodel}

\begin{figure}
\plottwo{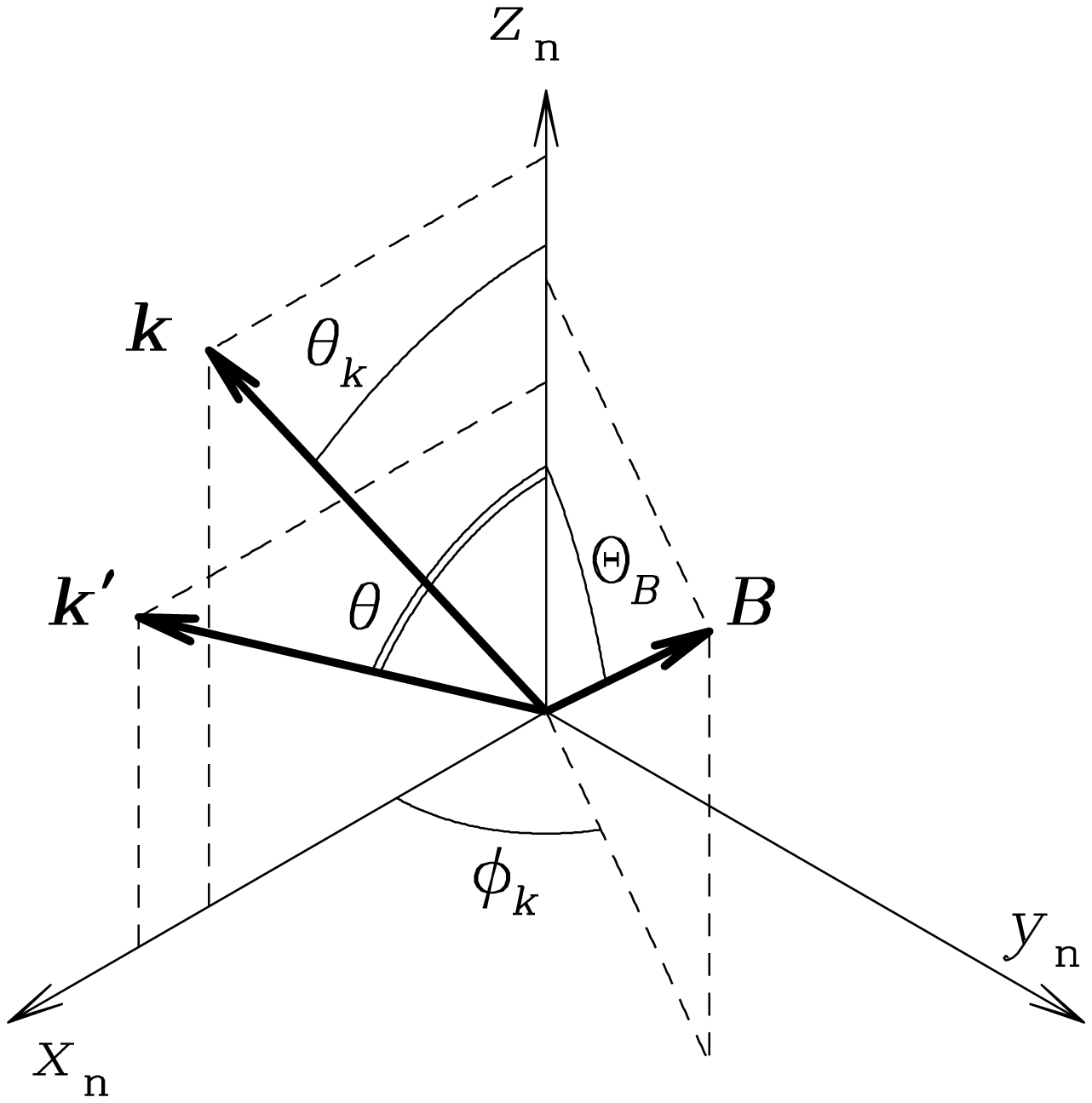}{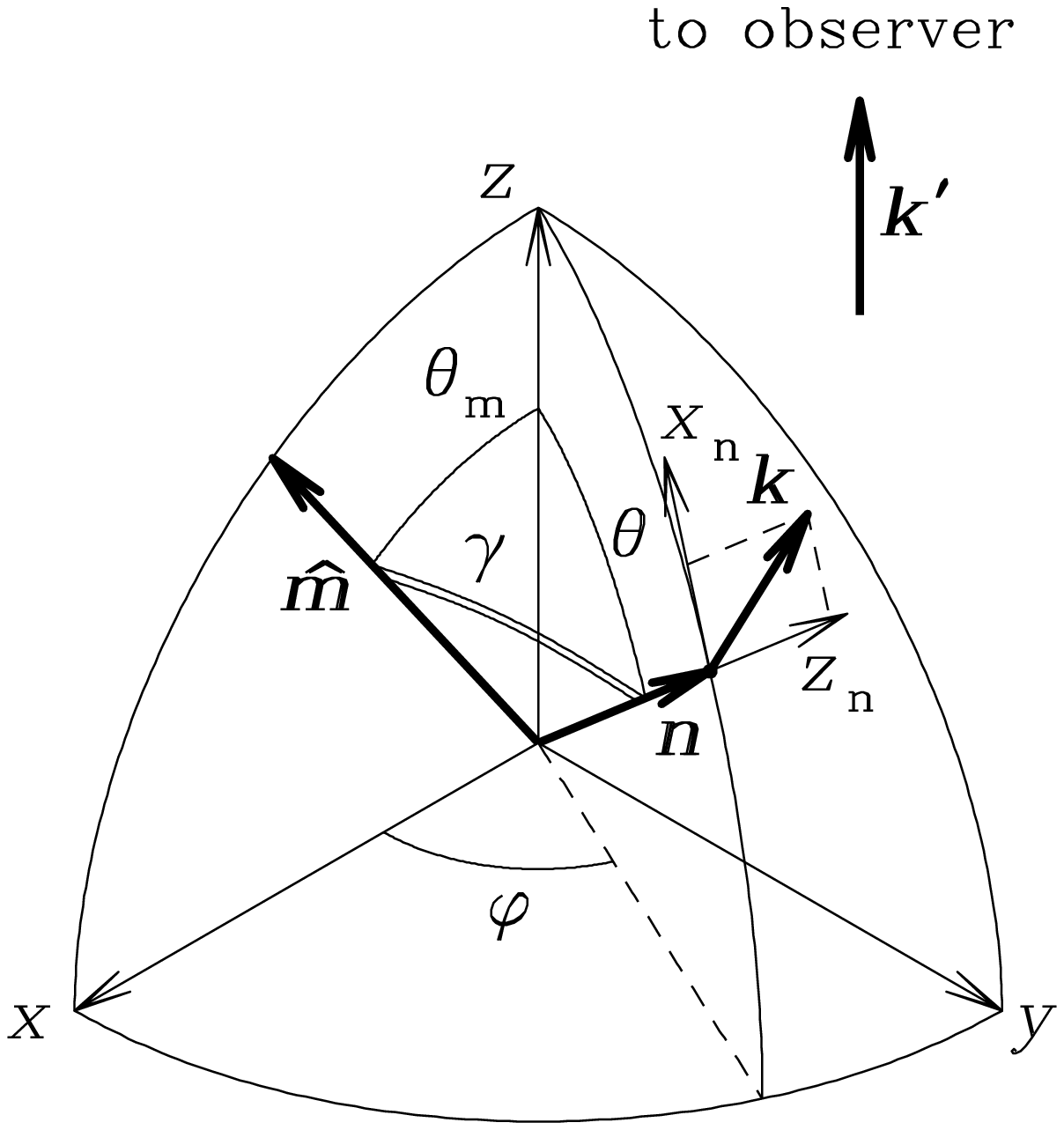}
\caption{
Coordinate axes and angles used to describe the atmosphere model.
Two Cartesian coordinate frames, $(x_\mathrm{n}y_\mathrm{n}z_\mathrm{n})$
and $(xyz)$, are considered.
The $z_\mathrm{n}$-axis is directed along the surface normal $\vecn$.
$\vecB$ is the magnetic field, and $\ThetaB$ and $\phik$ are its polar and
azimuthal angles in the $(x_\mathrm{n}y_\mathrm{n}z_\mathrm{n})$-frame.
$\veck$ is the photon wave-vector at the surface
and lies in the $(x_\mathrm{n}z_\mathrm{n})$-plane,
$\thetak$ is the angle between $\veck$ and $\vecn$,
and $\veckp$ is the photon wave-vector at infinity and is directed along $z$
($\veck=\veckp$ in the absence of gravitational light-bending).
$\theta$ and $\varphi$ are the polar and azimuthal angles of the surface
point at the stellar surface in the $(xyz)$-frame
(clearly, the angle between $\vecn$ and $\veckp$ equals $\theta$).
For a magnetic dipole, $\vecm$ is the unit vector along the magnetic 
axis, which lies in the
$(xz)$-plane, $\thetam$ is the angle between $\veckp$ and $\vecm$,
and $\gamma$ is the magnetic colatitude of the surface point.
\label{fig:angles}
}
\end{figure}

Thermal radiation from the surface of a NS is mediated by the stellar
atmosphere.
The model for emission through the atmosphere is constructed using a
grid in Thomson depth $\tau$, photon energy $E$, and photon propagation
direction $(\thetak,\phik)$, where $\thetak$ is the angle between the
photon wave-vector $\veck$ and the surface normal $\vecn$ and $\phik$
is the azimuthal angle between $\veck$ and the magnetic field $\vecB$
(see Fig.~\ref{fig:angles}).
Grid intervals are equally spaced logarithmically for depth
$10^{-5}\lesssim\tau\lesssim 10^3$ and energy $0.01\lesssim E\lesssim 10$~keV
and spaced every $5^\circ$ for $\thetak$ and $10^\circ$ for $\phik$
(extra grid points are included around $\thetak=\ThetaB$,
where $\ThetaB$ is the angle between $\vecn$ and $\vecB$);
6 grid points are used per decade in $\tau$, and 50 grid points are used
per decade in $E$.
Under typical conditions in NS atmospheres with $B\gtrsim 10^{12}$~G,
radiation propagates in two polarization modes
(see, e.g., \citealt{meszaros92});
therefore, the radiative transfer equations for the two coupled
modes are solved to determine the emission properties of
a magnetic atmosphere.
The self-consistency of the atmosphere model is determined by requiring
that the deviations (at each Thomson depth) from radiative equilibrium
and constant total flux are $\ll 1\%$ and $\lesssim 0.5\%$, respectively
(see \citealt{holai01,hoetal03,potekhinetal04}, and
references therein for details on the construction of the atmosphere models).
The atmosphere models mainly depend on three parameters:
the effective temperature $\Teff$ and the magnetic field strength
$B$ and inclination $\ThetaB$
(see Fig.~\ref{fig:angles}).
The atmosphere models also have a dependence, 
through hydrostatic balance,
on the surface gravity $g=(1+\zg)\,GM/R^2\approx
1.328\times10^{14}\,(1+\zg)\,(M/M_\odot)\,(R/10\mbox{~km})^{-2}$ cm~s$^{-2}$,
where  $\zg$ is the gravitational redshift, $1+\zg=(1-\rg/R)^{-1/2}$,
and $\rg=2GM/c^2\approx2.95\,(M/M_\odot)$~km is the gravitational radius.
Thus the atmosphere model depends on the NS mass $M$ and radius $R$;
however, the resulting spectra do not vary significantly for different
values of $g$ around $2\times 10^{14}$~cm~s$^{-2}$
(\citealt{pavlovetal95}; see Figs.~\ref{fig:sp12}, \ref{fig:sp712}, and
\ref{fig:sp13}).

The spectra from models constructed as discussed above only describe emission
from either a local patch of the stellar surface with the particular
$\Teff$ and $\vecB$ or a star with a uniform temperature and radial magnetic
field of uniform strength at the surface.
By taking into account surface magnetic field and temperature
distributions, we can construct more physical models of emission
from NSs;
however these spectra from the whole NS surface are necessarily
model-dependent,
as the $\vecB$ and $T$ distributions are generally unknown (see, e.g., 
\citealt{zavlinetal95a,zaneetal01,holai04,perezetal06,zaneturolla06,ho07};
see also
\citealt*{geppertetal06,aguileraetal07,potekhinetal07,reiseneggeretal07},
and references therein, for recent work on surface $B$ and $T$ distributions).
Therefore, we provide results from two sets of models: one set with a
single $\vecB$ and $\Teff$, as has been done previously
for NS atmospheres with fully ionized hydrogen
(see \nsa; \citealt{pavlovetal95}), and a set which is constructed with
$\vecB$ and $\Teff$ varying across the surface.
The latter is built by dividing the surface into regions with different
$\vecB$ and $\Teff$.
Relatively simple distributions are adopted (see Tables~\ref{tab:nssurf}
and \ref{tab:nsatm2}).
In particular, we assume a dipolar magnetic field, after accounting for
the effect of General Relativity, such that the surface distribution of
$\vecB$ is given by
\be
 \vecB = (\Bpole/2) \big[(2+f)(\vecn\cdot\vecm)\vecn-f\vecm\big]
\label{eq:dipole}
\ee
(\citealt{ginzburgozernoy65}; see also \citealt{pavlovzavlin00}),
where $\Bpole$ is the field strength at the magnetic pole,
$f=f(M/R)>1$ accounts for spatial curvature ($f=1$ in planar geometry),
and $\vecm$ is the direction of the magnetic moment.
The surface temperature distribution is calculated using the results of
\citet{potekhinetal03} for the ``canonical'' NS mass $M=1.4 M_\odot$ and 
the radius $R=12$~km, corresponding to moderately stiff NS EOSs
(see, e.g., \citealt*{haenseletal07}).
In order to minimize model dependence, we assume the $\Teff$-distribution
of an iron heat-blanketing envelope.  This assumption does not change our
results since, for any chemical composition of the envelope, the dependence
of $\Teff$ on the magnetic colatitude $\gamma$ is similar to that given
in \citet{greensteinhartke83} (see \citealt*{potekhinetal03}).

\begin{table}
\caption{Magnetic Field Distribution of Neutron Star Surface
 \label{tab:nssurf}}
\begin{tabular}{ccccc}
\hline
$\gamma$ & $\ThetaB$ & \multicolumn{2}{c}{$B$} \\
(deg) & (deg) & \multicolumn{2}{c}{($10^{12}$ G)} \\
\hline
  &   & case 1 &  case 2 \\
0 & 0 & 10 & 1.82 \\
30 & 20 & 9 & 1.65 \\
60 & 45 & 7 & 1.26 \\
90 & 90 & 5.5 & 1 \\
\hline
\end{tabular}
\end{table}

\begin{table}
\small
\caption{Effective Temperatures $\Teff/(10^5\mbox{ K})$ of Neutron Star
Atmosphere Models \label{tab:nsatm2}}
\begin{tabular}{c c c c c}
\hline
$\log\Teff^\mathrm{NS}$ & \multicolumn{4}{c}{$\ThetaB$} \\
& 0$^\circ$ & 20$^\circ$ & 45$^\circ$ & 90$^\circ$ \\
\hline
& \multicolumn{4}{c}{$B$ ($10^{12}$ G)} \\
& 1.82 & 1.65 & 1.26 & 1 \\ \\
 5.5 & 3.8 & 3.7 & 3.2 & 0.55\tablenotemark{a} \\
 5.6 & 4.7 & 4.6 & 4.0 & 0.82\tablenotemark{a} \\
 5.7 & 6.0 & 5.8 & 5.1 & 1.2\tablenotemark{a} \\
 5.8 & 7.5 & 7.3 & 6.4 & 1.8\tablenotemark{a} \\
 5.9 & 9.4 & 9.2 & 8.0 & 2.66 \\
 6.0 & 12 & 11.5 & 10 & 3.9 \\
 6.1 & 15 & 14 & 13 & 5.66 \\
 6.2 & 19 & 18 & 16 & 8.1 \\
 6.3 & 23.3 & 22.8 & 20 & 11.6 \\
 6.4 & 29 & 28 & 25 & 16 \\
 6.5 & 36 & 35 & 32 & 22.3 \\
 6.6 & 45 & 44 & 40 & 30.4 \\
 6.7 & 56 & 55 & 50 & 40.7 \\
 6.8 & 70 & 69 & 63 & 54 \\
\\
& \multicolumn{4}{c}{$B$ ($10^{12}$ G)} \\
& 10 & 9 & 7 & 5.5 \\ \\
 5.5 & 3.75 & 3.66 & 3.2 & 0.27\tablenotemark{a} \\
 5.6 & 4.7 & 4.6 & 4.04 & 0.4\tablenotemark{a} \\
 5.7 & 5.94 & 5.8 & 5.1 & 0.59\tablenotemark{a} \\
 5.8 & 7.5 & 7.3 & 6.4 & 0.86\tablenotemark{a} \\
 5.9 & 9.47 & 9.2 & 8.0 & 1.3\tablenotemark{a} \\
 6.0 & 12 & 11.6 & 10 & 1.9\tablenotemark{a} \\
 6.1 & 15 & 14.6 & 12.7 & 2.9 \\
 6.2 & 19 & 18.3 & 16 & 4.4 \\
 6.3 & 23.6 & 23 & 20 & 6.5 \\
 6.4 & 29.6 & 29 & 25.5 & 9.5 \\
 6.5 & 37 & 36 & 32 & 13.8 \\
 6.6 & 46.6 & 46 & 40 & 19.8 \\
 6.7 & 58.4 & 57 & 51 & 28 \\
 6.8 & 73 & 72 & 64 & 39 \\
\hline
\end{tabular}
\tablecomments{Models assume $g=1.6\times 10^{14}$ cm s$^{-2}$.
$^{a}$Spectra for these temperatures use blackbodies with
$T=\Teff$. \label{tabnote:bb}}
\end{table}

Emission from any point along a circle at a fixed $\gamma$ is given by the
atmosphere model for that $\gamma$.
Using equation~(\ref{eq:dipole}) and the coordinate transformation
\begin{eqnarray}
\hat{\bm{x}}_\mathrm{n} &=& - \cos\varphi\cos\theta\,\hat{\bm{x}}
 - \sin\varphi\cos\theta\,\hat{\bm{y}} + \sin\theta\,\hat{\bm{z}} \nonumber \\
\hat{\bm{y}}_\mathrm{n} &=& - \sin\varphi\,\hat{\bm{x}}
 - \cos\varphi\,\hat{\bm{y}} \\
\hat{\bm{z}}_\mathrm{n} &=& \cos\varphi\sin\theta\,\hat{\bm{x}}
 + \sin\varphi\sin\theta\,\hat{\bm{y}} + \cos\theta\,\hat{\bm{z}}, \nonumber
\end{eqnarray}
where $\theta$ and $\varphi$ are the polar and azimuthal angles of
$\vecn$ with respect to the line of sight (see Fig.~\ref{fig:angles})
and ($\hat{\bm{x}},\hat{\bm{y}},\hat{\bm{z}}$) and
($\hat{\bm{x}}_\mathrm{n},\hat{\bm{y}}_\mathrm{n},\hat{\bm{z}}_\mathrm{n}$)
are the unit coordinate vectors for frames ($x,y,z$) and
($x_\mathrm{n},y_\mathrm{n},z_\mathrm{n}$), respectively,
 one finds that
\begin{eqnarray}
\cos\gamma &=& \cos\varphi\sin\theta\sin\thetam + \cos\theta\cos\thetam, \\
\sin\phik &=& \sin\varphi\sin\thetam/\sin\gamma, \\
B &=& \Bpole\sqrt{\cos^2\gamma + (f/2)^2 \sin^2\gamma}, \label{eq:dipole2} \\
\tan\ThetaB &=& (f/2)\tan\gamma,
\label{eq:sinphik}
\end{eqnarray}
where $\thetam$ is the angle between $\vecm$ and $\veckp$ and
$\phik=\varphi$ in the special case $\gamma=0$.
The photon wave-vector at infinity $\veckp$ differs from $\veck$ at the
surface due to gravitational redshift and light-bending
\citep*{pechenicketal83,page95,pavlovzavlin00}.
The latter is taken into account by making use of the approximation from
\cite{beloborodov02} (see also \citealt*{zavlinetal95b}),
\be
1 - \cos\theta = (1-\cos\thetak)/(1-\rg/R).
\label{eq:lightbend}
\ee
The spectrum for the entire surface is then computed by summing over
the emission from different regions,
\be
F_E = \fluxnorm \int_0^{2\pi}\mathrm{d}\varphi\int_0^{\pi/2}
 \sin\thetak \mathrm{d}\thetak
 I_E(\thetak,\phik;B,\ThetaB),
\label{eq:flux1}
\ee
where
$I_E$ is the specific intensity.
Note that $E$ is the unredshifted photon energy.  The explicit redshifting
of the photon energy and flux spectrum is not done at this stage, though
relativistic effects are taken into account [see, e.g.,
eqs.~(\ref{eq:dipole2}) and (\ref{eq:lightbend})]; redshifting is
done in the \xspec\ fitting code (see \S~\ref{sec:results}).
At this point in the model calculation, the flux normalization
$\fluxnorm$ [$=\fluxnorm(M,R,d)$, where $d$ is the distance to the source]
is taken to be unity (see \S~\ref{sec:results} for discussion of its
model-dependence).

We calculate $I_E$ for four $\gamma$ values (see Table~\ref{tab:nssurf})
and perform the integration in equation~(\ref{eq:flux1}) by interpolating
between the calculated values.
For $\Teff(\gamma=90^\circ)<2\times 10^5$~K, we use blackbody spectra at
$T=\Teff$ (see Table~\ref{tab:nsatm2}); the spectral contributions at
these temperatures contribute little to the total X-ray spectra [since
$\Teff^\mathrm{NS}\gg\Teff(\gamma=90^\circ)$; replacing the blackbody
spectrum with zero values yields no appreciable difference in the resulting
integrated spectrum], which is dominated by emission from the hotter
regions of the NS surface.
Strong absorption features, such as the proton cyclotron line at
$E_{B\mathrm{p}}=\hbar eB/\mpr c$, are broadened due to the variation of
$B$ with $\gamma$.
In order to reproduce this broadening in our interpolation, we first
remap our calculated $I_E$ as a function of $E/E_{B\mathrm{p}}$:
$I_E(\thetak,\phik;B,\ThetaB)
          \equiv\tilde{I}(E/E_{B\mathrm{p}},\thetak,\phik;\gamma)$.
We then interpolate $\tilde{I}$ in $\thetak$, $\phik$, and $B(\gamma)$
for every fixed $E/E_{B\mathrm{p}}$ and substitute the resulting $I_E$
into equation~(\ref{eq:flux1}).

\section{Results} \label{sec:results}

\begin{figure}
\plotone{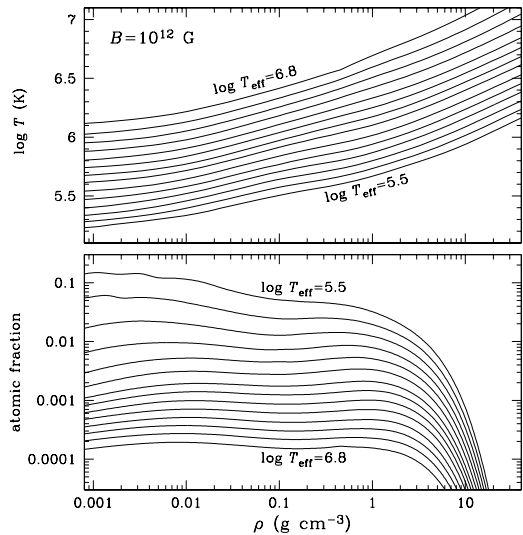}
\caption{
Temperature and abundance profiles at various $\Teff$
of a partially ionized hydrogen atmosphere with
$B=10^{12}$~G, $\ThetaB=0^\circ$, and $g=2.4\times 10^{14}$~cm~s$^{-2}$.
\label{fig:t12}
}
\end{figure}

\begin{figure}
\plotone{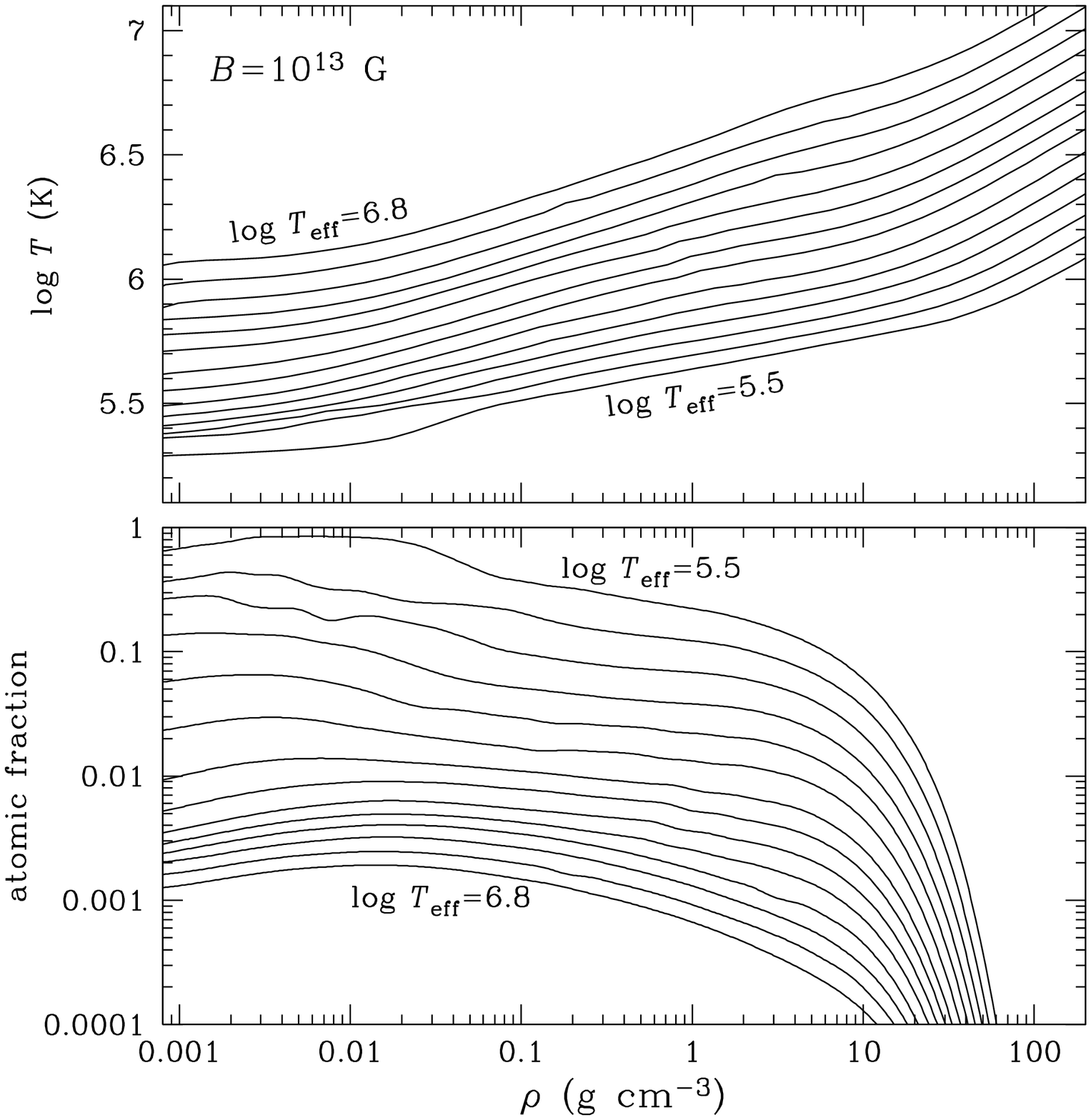}
\caption{
Temperature and abundance profiles at various $\Teff$
of a partially ionized hydrogen atmosphere with
$B=10^{13}$~G, $\ThetaB=0^\circ$, and $g=2.4\times 10^{14}$~cm~s$^{-2}$.
\label{fig:t13}
}
\end{figure}

For the first set of models (with uniform $\vecB$ and $\Teff$),
we consider $g=1.6$ and $2.4\times 10^{14}$~cm~s$^{-2}$;
for a NS with $M=1.4 M_\odot$, this corresponds to $R=12$~km and 10~km,
respectively.
The magnetic field is $B=10^{12}$, $1.26\times 10^{12}$, $2\times 10^{12}$,
$4\times 10^{12}$, $7\times 10^{12}$, $10^{13}$, or $2\times 10^{13}$~G
and $\ThetaB=0^\circ$.
The effective temperatures span the range $\log\Teff=5.5$--$6.8$
($5.6$--$6.8$ for $B=2\times 10^{13}$~G) with the
temperature interval between each model $\Delta\log\Teff\approx 0.1$.
The temperature and abundance profiles for the atmosphere models with
$g=2.4\times 10^{14}$~cm~s$^{-2}$ and $B=10^{12}$~G and $10^{13}$~G
are shown in Figures~\ref{fig:t12} and \ref{fig:t13}, respectively.
The atomic fraction is the number of hydrogen atoms with non-destroyed
energy levels divided by the total number of protons
\citep{potekhinchabrier03}.
Here we only account for the ground-state atoms,
which substantially reduces the computational work.
This approximation is justified
because the fraction of atoms in excited states
is small in most of the considered temperature profiles:
it does not exceed 
a few percent even when the abundance of ground-state atoms
reaches tens of percent.
The dependence of the atomic fraction on temperature and magnetic field
is clear: lower temperatures or higher magnetic fields increase the
abundance of bound species. The dependence on density is more complex:
an increase in the atomic fraction with growing density
(recombination according to the modified Saha equation) 
competes with the decrease due to pressure ionization,
which ultimately turns to complete ionization at high $\rho$.

\begin{figure}
\plotone{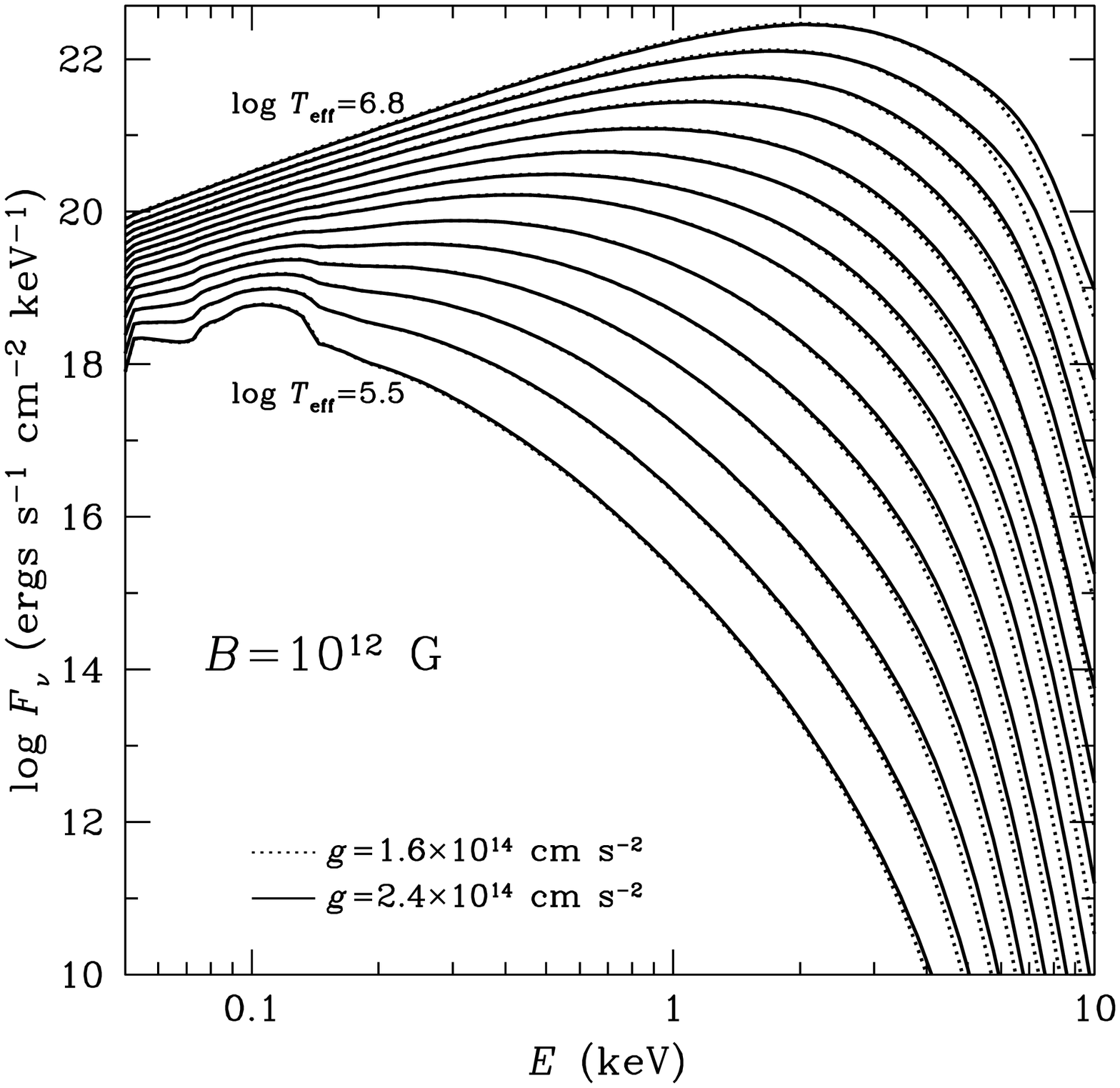}
\caption{
Spectra at various $\Teff$ of a partially ionized hydrogen atmosphere
with $B=10^{12}$~G and $\ThetaB=0^\circ$.
Dotted and solid lines are for $g=1.6$ and $2.4\times 10^{14}$~cm~s$^{-2}$,
respectively.
\label{fig:sp12}
}
\end{figure}

\begin{figure}
\plotone{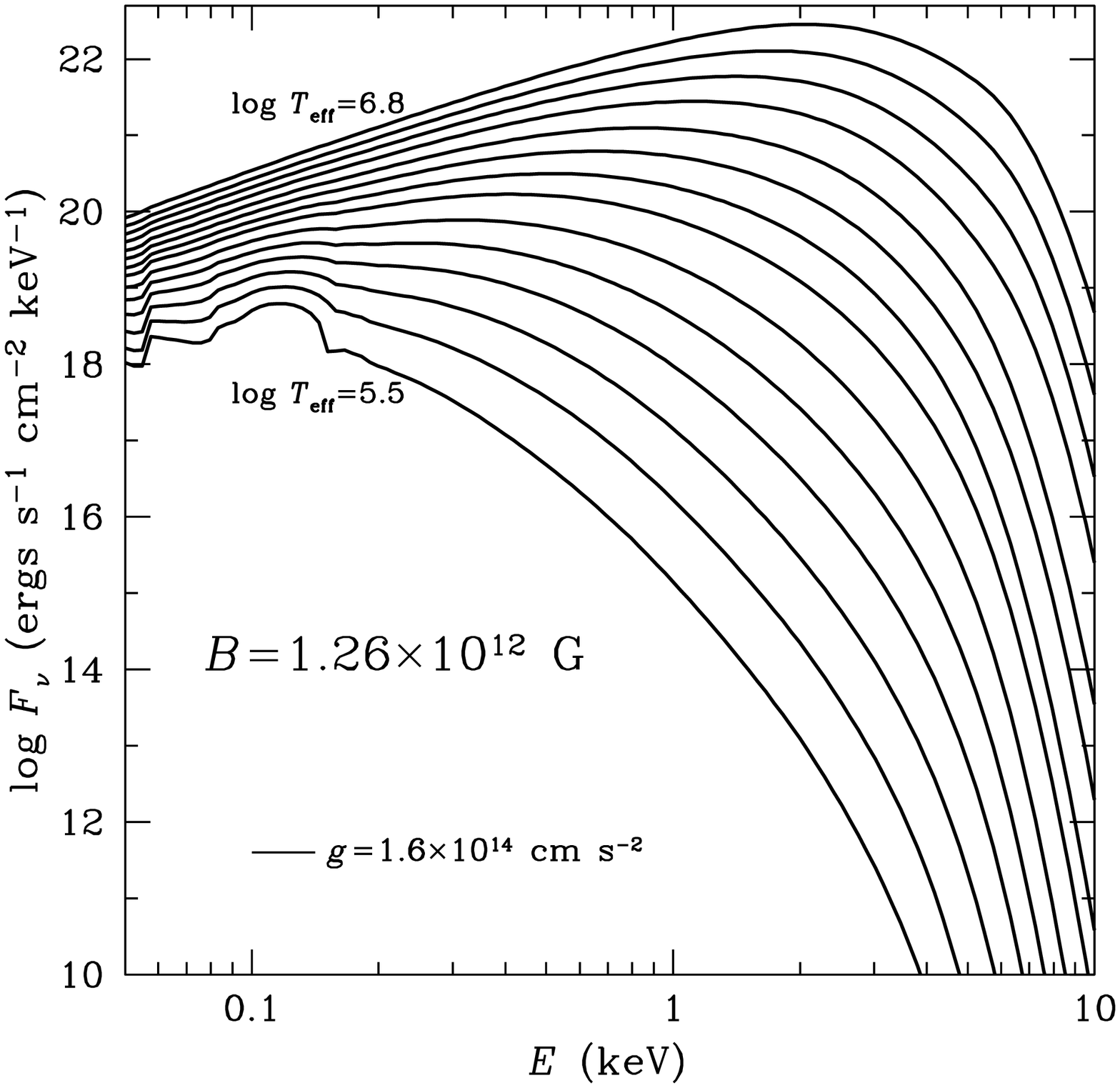}
\caption{
Spectra at various $\Teff$ of a partially ionized hydrogen atmosphere
with $B=1.26\times 10^{12}$~G, $\ThetaB=0^\circ$, and
$g=1.6\times 10^{14}$~cm~s$^{-2}$.
\label{fig:sp12612}
}
\end{figure}

\begin{figure}
\plotone{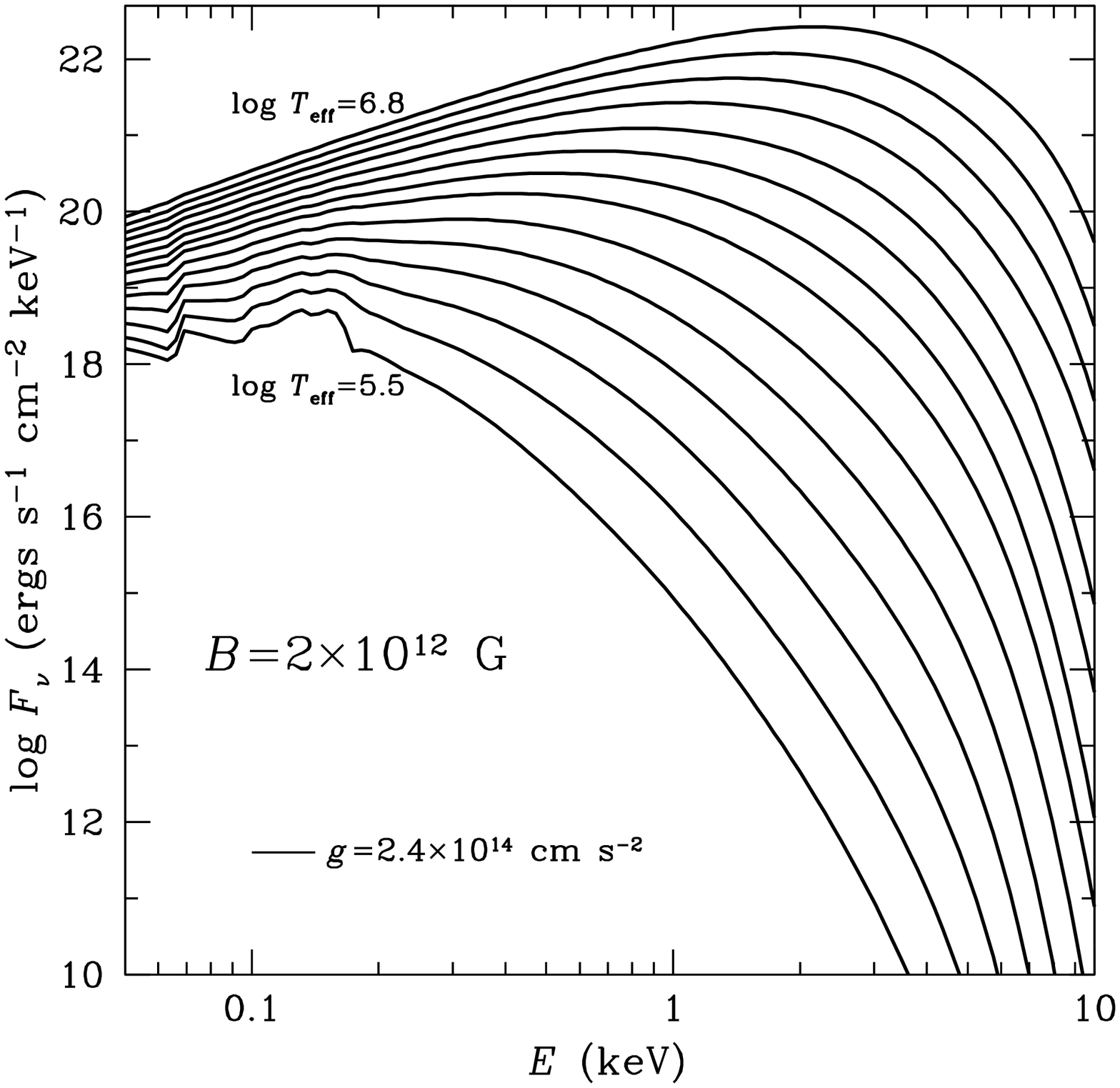}
\caption{
Spectra at various $\Teff$ of a partially ionized hydrogen atmosphere
with $B=2\times 10^{12}$~G, $\ThetaB=0^\circ$, and
$g=2.4\times 10^{14}$~cm~s$^{-2}$.
\label{fig:sp212}
}
\end{figure}

\begin{figure}
\plotone{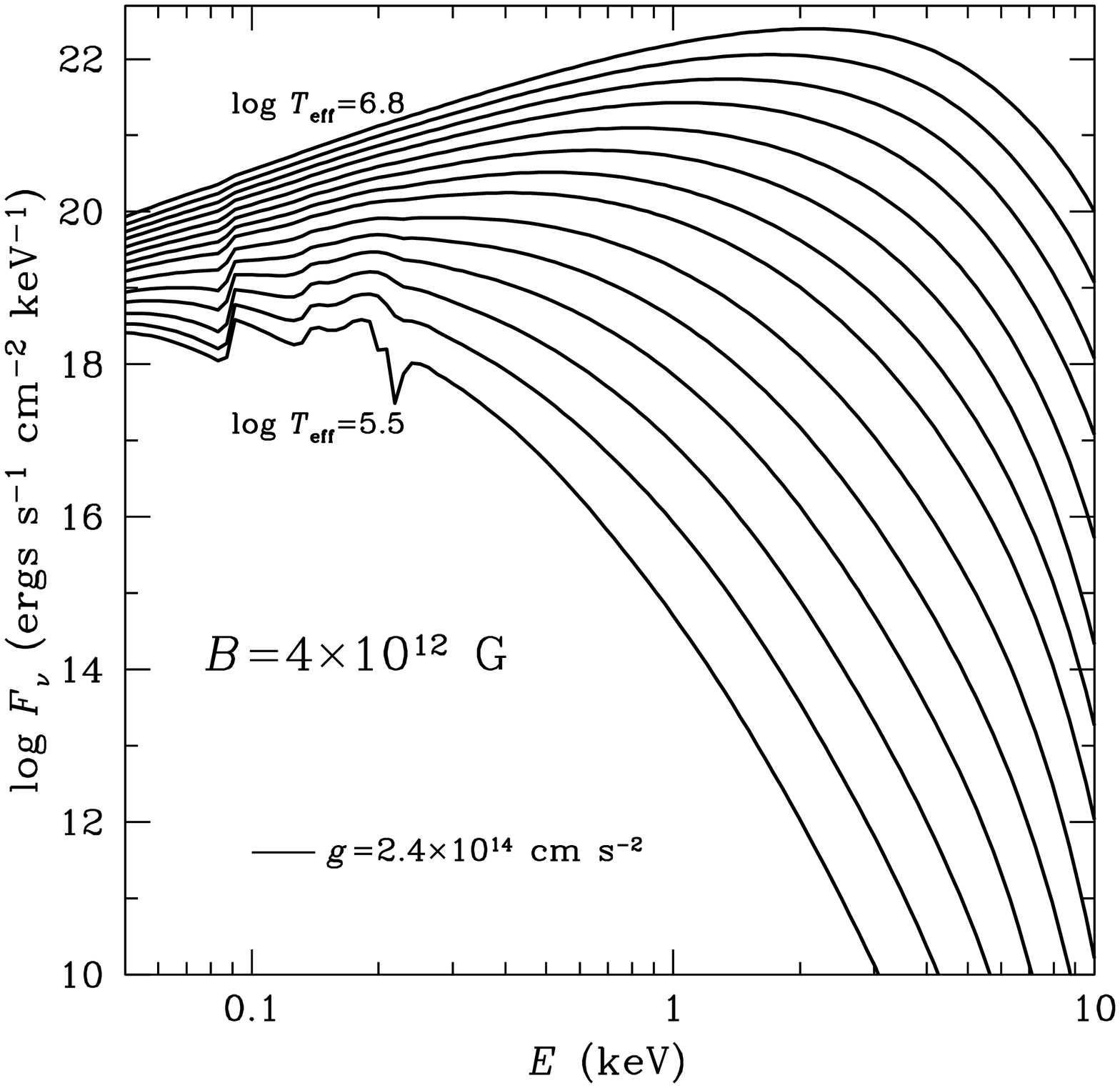}
\caption{
Spectra at various $\Teff$ of a partially ionized hydrogen atmosphere
with $B=4\times 10^{12}$~G, $\ThetaB=0^\circ$, and
$g=2.4\times 10^{14}$~cm~s$^{-2}$.
\label{fig:sp412}
}
\end{figure}

\begin{figure}
\plotone{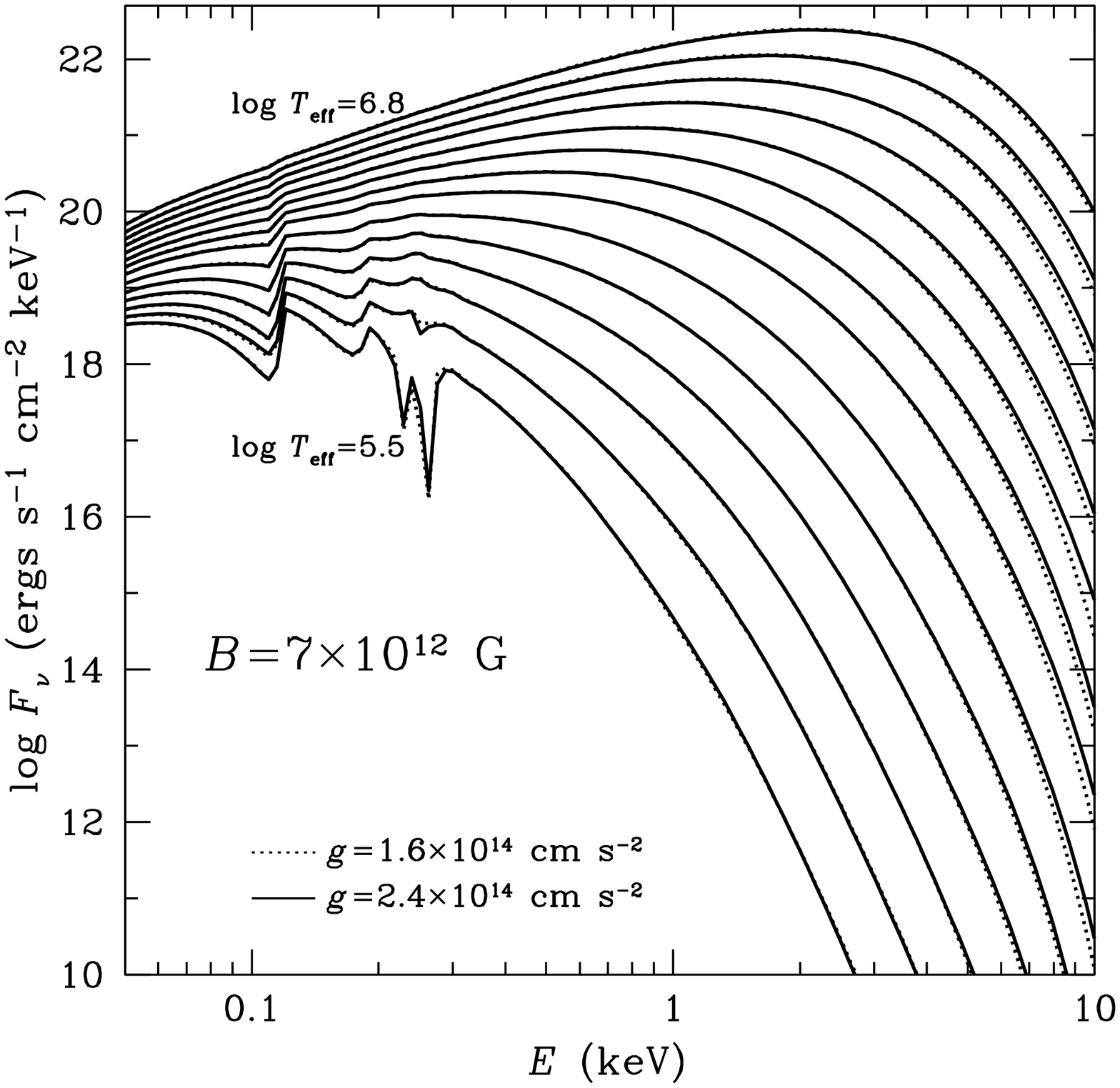}
\caption{
Spectra at various $\Teff$ of a partially ionized hydrogen atmosphere
with $B=7\times 10^{12}$~G and $\ThetaB=0^\circ$.
Dotted and solid lines are for $g=1.6$ and $2.4\times 10^{14}$~cm~s$^{-2}$,
respectively.
\label{fig:sp712}
}
\end{figure}

\begin{figure}
\plotone{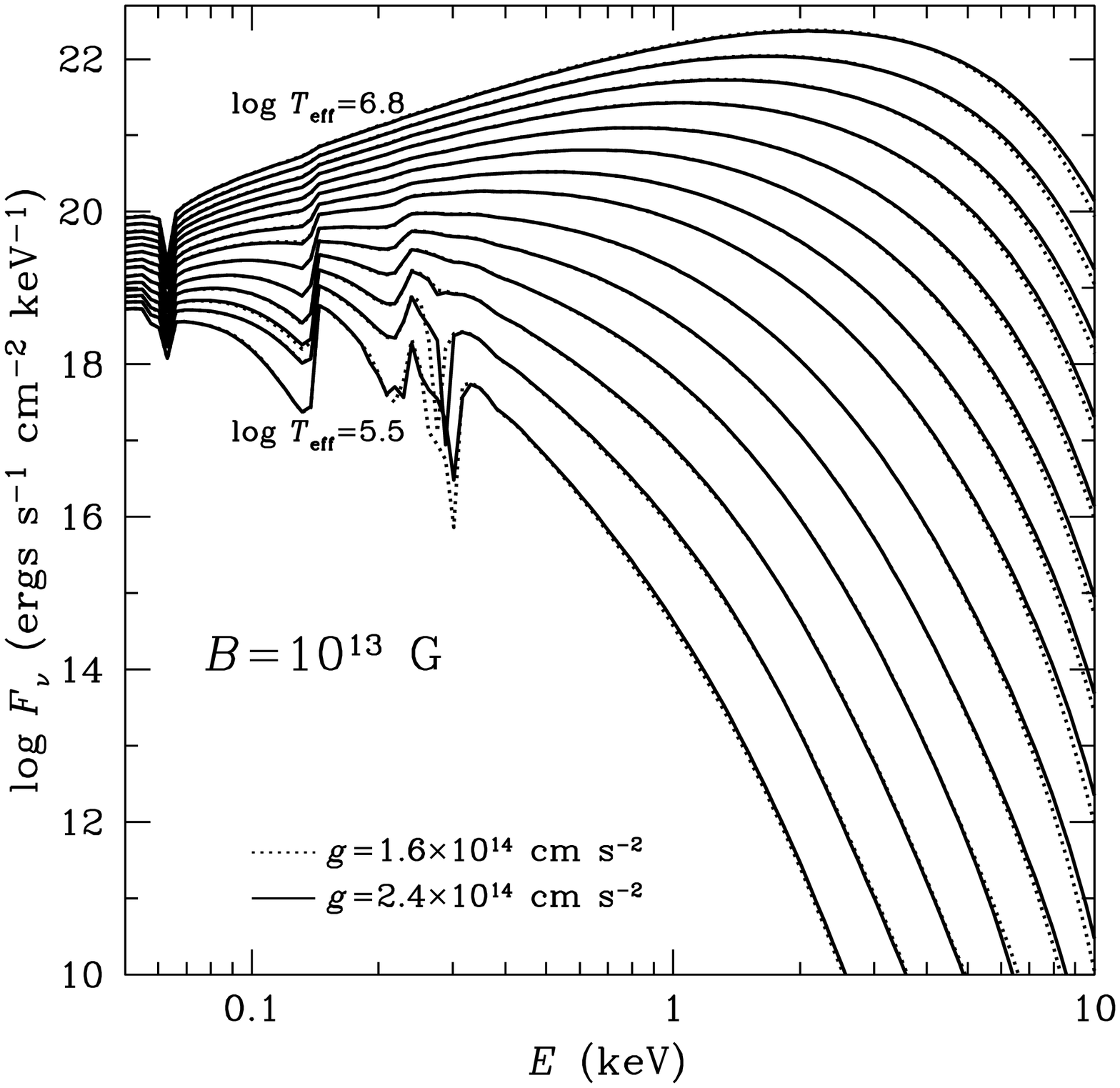}
\caption{
Spectra at various $\Teff$ of a partially ionized hydrogen atmosphere
with $B=10^{13}$~G and $\ThetaB=0^\circ$.
Dotted and solid lines are for $g=1.6$ and $2.4\times 10^{14}$~cm~s$^{-2}$,
respectively.
\label{fig:sp13}
}
\end{figure}

\begin{figure}
\plotone{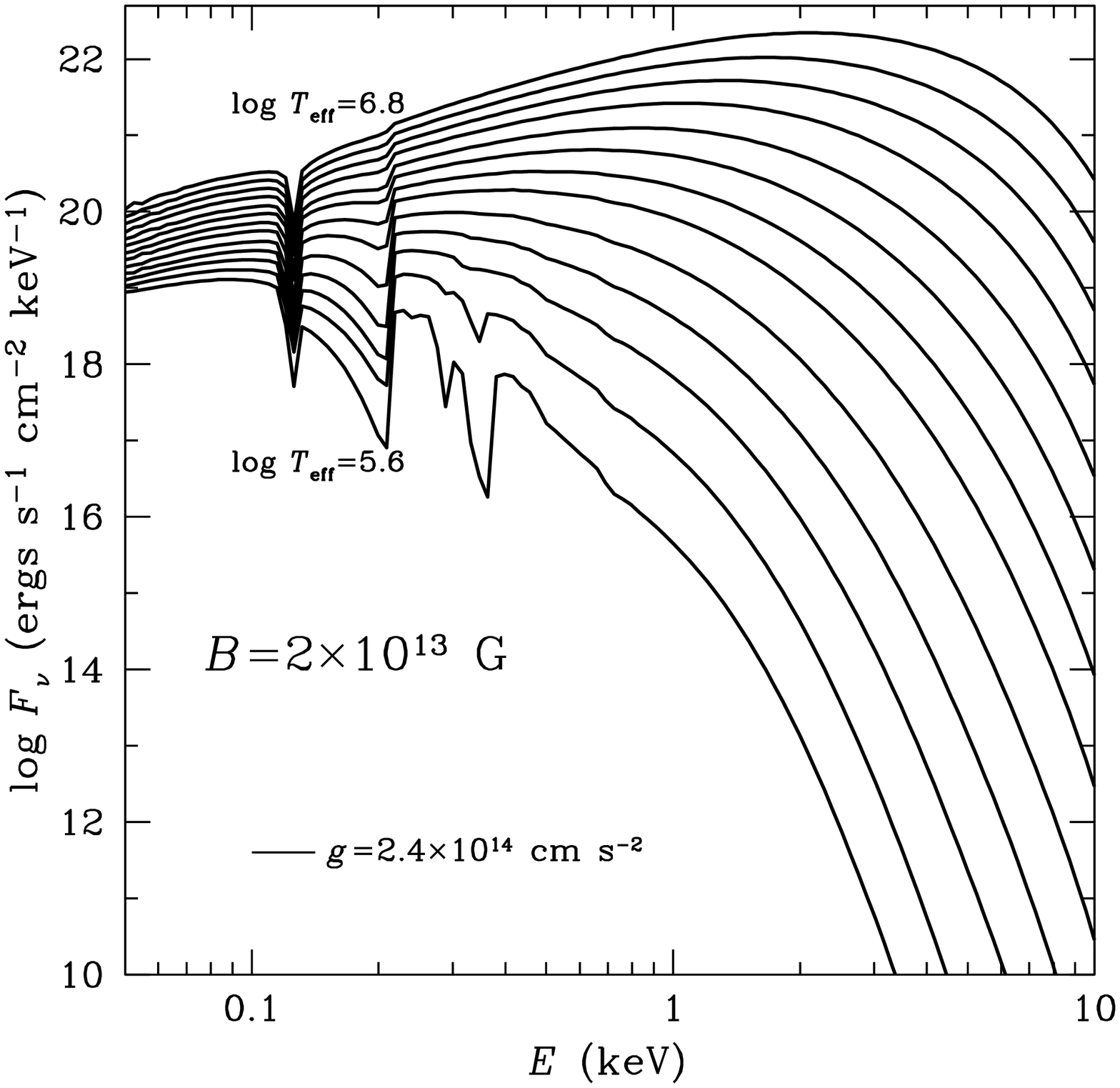}
\caption{
Spectra at various $\Teff$ of a partially ionized hydrogen atmosphere
with $B=2\times 10^{13}$~G, $\ThetaB=0^\circ$, and
$g=2.4\times 10^{14}$~cm~s$^{-2}$.
\label{fig:sp213}
}
\end{figure}

Figures~\ref{fig:sp12}--\ref{fig:sp213} show the (unredshifted) spectra
for $B=10^{12}$, $1.26\times 10^{12}$, $2\times 10^{12}$, $4\times 10^{12}$,
$7\times 10^{12}$, $10^{13}$, and $2\times 10^{13}$~G.
The most prominent spectral features are due to 
the proton cyclotron line at
$E_{B\mathrm{p}}=0.063\,(B/10^{13}\mbox{ G})$~keV,
the $s=0\rightarrow 1$ transition at
$E=0.051$~keV for $B=10^{12}$~G and $0.14$~keV for $10^{13}$~G,
the $s=0\rightarrow 2$ transition at
$0.075$~keV for $10^{12}$~G and $0.23$~keV for $10^{13}$~G,
and the peak of the bound-free transition at
$0.16$~keV for $10^{12}$~G and $0.31$~keV for $10^{13}$~G.
Here $s$ is the quantum number that measures transverse atomic excitations
and corresponds to the projection of the angular momentum onto the magnetic
field lines, whereas the longitudinal and Landau quantum numbers equal zero
for the bound states involved in these transitions (see \citealt{potekhin94},
for a detailed description of the quantum numbers of a moving hydrogen atom).
All the spectral features due to atomic transitions are substantially
broadened because of the ``motional Stark effect''
(see, e.g., \citealt{potekhinpavlov97}, and references therein).
This ``magnetic broadening'' becomes stronger with increasing $T$ and
is another reason, in addition to the decrease in the neutral fraction,
for the disappearance of the features from the spectra at higher $\Teff$.

\begin{figure}
\plotone{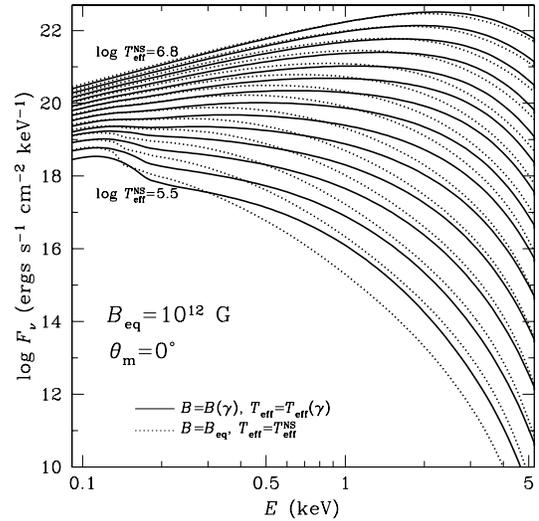}
\caption{
Spectra at various total NS $\Teff^\mathrm{NS}$ of a partially ionized
hydrogen atmosphere covering a NS with $R=12$~km and $M=1.4 M_\odot$,
surface distributions of $\vecB$ and $\Teff$ according to the magnetic
dipole model (see Tables~\ref{tab:nssurf} and \ref{tab:nsatm2};
$\Bpole=1.82\times 10^{12}$~G and $B=10^{12}$~G at the magnetic equator),
and $\thetam=0$. 
Dotted lines correspond to atmosphere spectra with a uniform temperature
($\Teff=\Teff^\mathrm{NS}$) and uniform radial magnetic field $B=10^{12}$~G.
\label{fig:sp12eq}
}
\end{figure}

\begin{figure}
\plotone{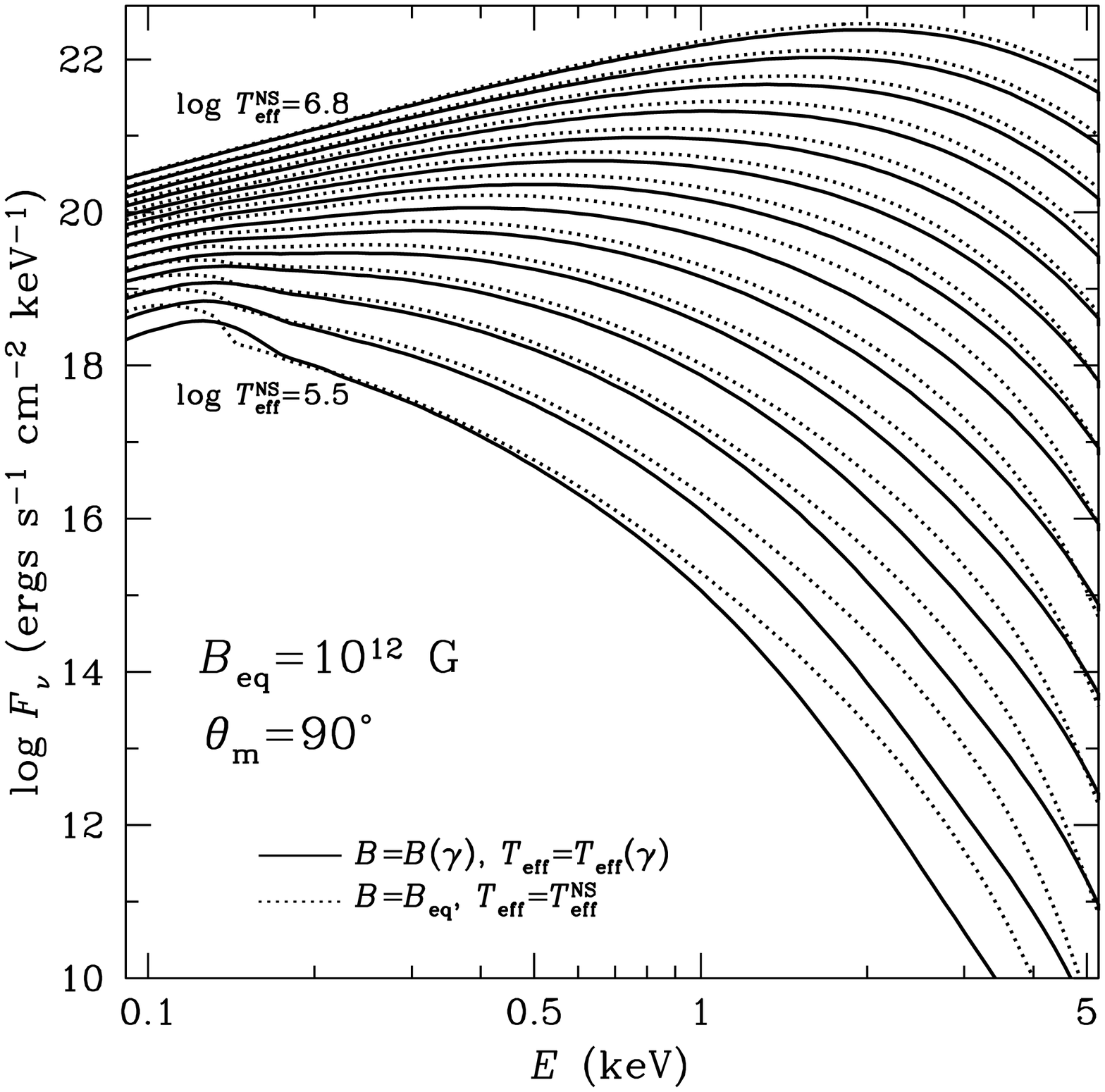}
\caption{
The same as in Fig.~\ref{fig:sp12eq} but with $\thetam=90^\circ$.
\label{fig:sp12eq90}
}
\end{figure}

\begin{figure}
\plotone{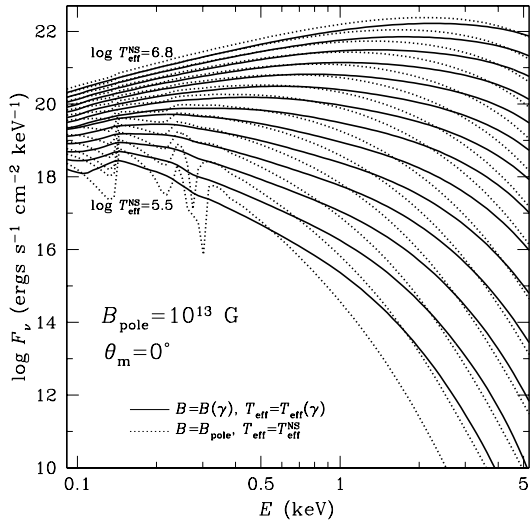}
\caption{
The same as in Fig.~\ref{fig:sp12eq} but with $\Bpole=10^{13}$~G and
$B=5.5\times 10^{12}$~G at the equator.
Dotted lines correspond to atmosphere spectra with a uniform radial
magnetic field $B=10^{13}$~G.
\label{fig:sp13pole0}
}
\end{figure}

\begin{figure}
\plotone{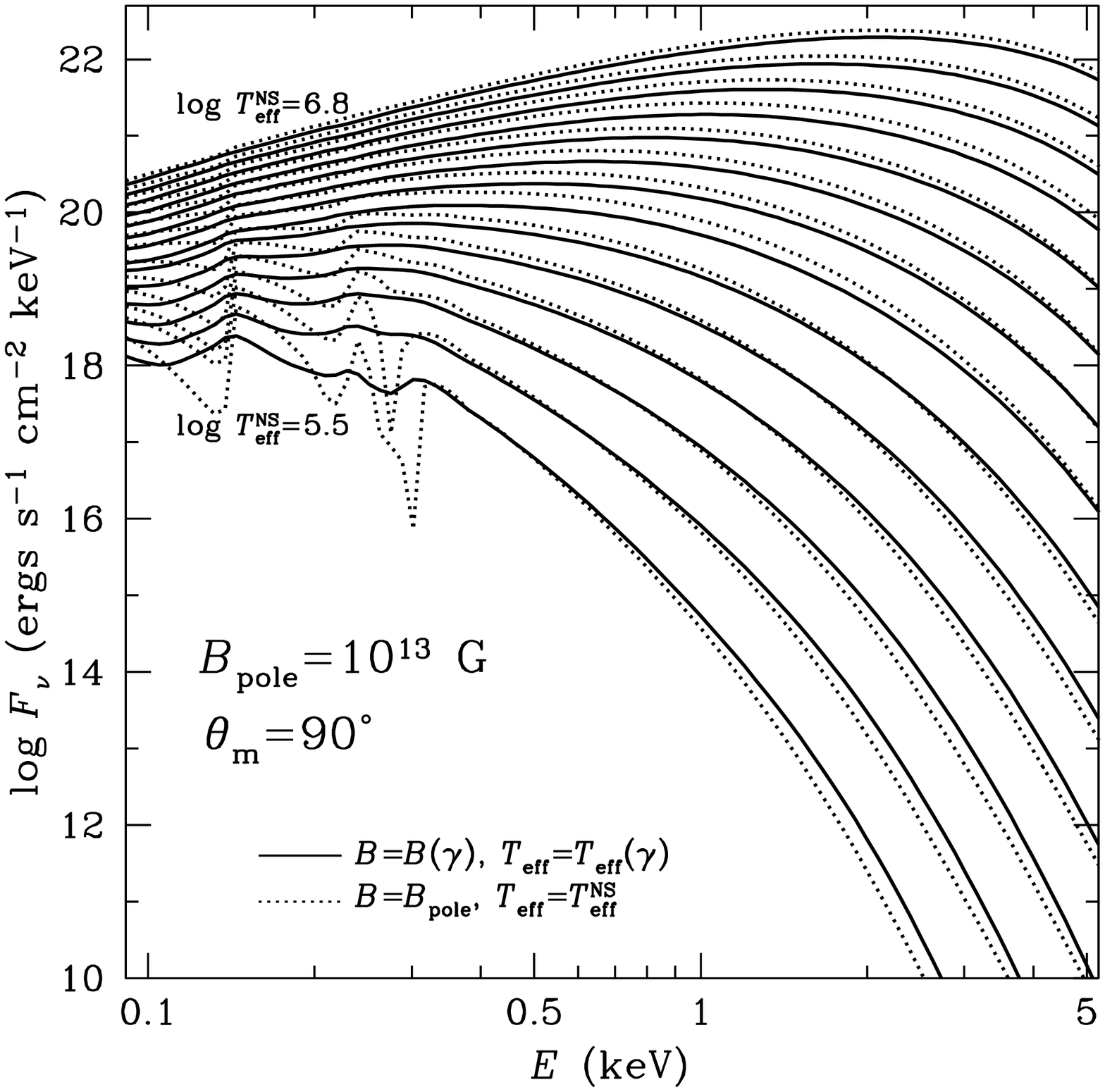}
\caption{
The same as in Fig.~\ref{fig:sp13pole0} but with $\thetam=90^\circ$.
}
\label{fig:sp13pole}
\end{figure}

For the second set of models, we take $g=1.6\times 10^{14}$~cm~s$^{-2}$.
The range of magnetic fields and effective temperatures $(B,\ThetaB,\Teff)$
of the models are given in Table~\ref{tab:nsatm2}.
These values correspond to a magnetic dipole model of
a NS with $R=12$~km and $M=1.4 M_\odot$, in agreement with 
the chosen $g$.
One subset of models has $B=10^{12}$~G at the magnetic equator,
while the other has $B=10^{13}$~G at the pole.
$\Teff^\mathrm{NS}$ is the mean effective temperature for the
whole NS, which corresponds to the total heat flux
from the surface (see, e.g., \citealt{potekhinetal03}).
For each $\Teff^\mathrm{NS}$,
the spectra are calculated as described in \S~\ref{sec:atmmodel} and
using equation~(\ref{eq:flux1}).
The resulting spectra are shown in
Figures~\ref{fig:sp12eq}--\ref{fig:sp13pole}
for the cases $\thetam=0$ and $90^\circ$.
For comparison, the dotted lines show the atmosphere spectra for a
uniform surface temperature and radial magnetic field
(see Figs.~\ref{fig:sp12} and \ref{fig:sp13}).
We see that the field distribution over the stellar surface
substantially smears the spectral features from atomic transitions.
This smearing is especially noticeable in Figures~\ref{fig:sp13pole0}
and \ref{fig:sp13pole}, where the atomic features are stronger
due to the higher atomic fractions.

Finally, we supply \xspec, under the model name \nsmax, with tables of
the spectra shown in Figures~\ref{fig:sp12}--\ref{fig:sp13pole},
as well as a code to interpolate within each table;
note that the model spectra with single ($\vecB,\Teff$)-values span the
photon energy range $0.05\lesssim E\le 10$~keV, while the model spectra
with ($\vecB,\Teff$)-distributions cover $0.09\lesssim E\lesssim 5$~keV.
The code first unredshifts the energy bins of the observed spectrum, then
obtains the fit spectrum by linear interpolation via a weighted average of
the nearest two model $\log\Teff$ and $E$, and finally redshifts the fit
spectrum by $(1+\zg)^{-1}$.  The code requires one switch parameter and
two fit parameters ($\log\Teff$ and $1+\zg$); \xspec\ automatically adds
a third fit parameter (normalization $\fluxnorm$).
The switch parameter indicates which table of model spectra to use:
the differences being due to the composition (only hydrogen at the present
time; see \S~\ref{sec:discussion}), $B$, $\ThetaB$, and $g$ for the first
set of models and the composition, $B$, $\thetam$, and $g$ for the second
set of models.
The normalization parameter $\fluxnorm$ is conventionally taken to be
equal to $(\Rinfty/d)^2/(1+\zg)^3$, where $\Rinfty = R\,(1+\zg)$,
and the same $R$ is used to calculate $\zg$; note that this prescription
implies the emission region is the entire visible surface of the NS.

\section{Summary} \label{sec:discussion}

We have constructed tables of model atmosphere spectra for neutron stars
(with magnetic fields $B=10^{12}$--$2\times 10^{13}$~G and effective
temperatures $\log\Teff=5.5$--$6.8$) and incorporated these tables into
\xspec\ (under the model name
\nsmax\footnote{http://heasarc.gsfc.nasa.gov/docs/xanadu/xspec/models/nsmax.html}).
These spectra are obtained using the most up-to-date equation of state
and opacities for a partially ionized hydrogen plasma, and therefore
they can describe emission from neutron stars with surface temperatures
$T\lesssim 10^6$~K, where the abundance of bound species is appreciable,
as well as neutron stars with $T> 10^6$~K.
Thus we go beyond the previous magnetic neutron star spectral models
provided in \xspec, which assume fully ionized hydrogen atmospheres.
Our implementation in \xspec\ allows easy updates to the database
of model spectra, so that tables of models with other magnetic field
strengths (e.g., other than the seven fields, $B=10^{12}$,
$1.26\times 10^{12}$, $2\times 10^{12}$, $4\times 10^{12}$, $7\times 10^{12}$,
$10^{13}$, and $2\times 10^{13}$~G, currently provided)
or other elements (e.g., carbon, oxygen, and neon; see \citealt{moriho07})
will be added as they become available;
the \xspec-user merely specifies a switch parameter to indicate which
set of models is to be used in the fitting.
We have also constructed tables of model spectra that account for
relativistic effects and dipolar magnetic field and temperature variations
on the surface of the neutron star.
These spectra are more realistic but also more model-dependent.
They show significant smearing of spectral features 
compared to the models that assume a uniform magnetic field.

\acknowledgements

We thank Keith Arnaud for assistance in incorporating \nsmax\ into \xspec\
and the anonymous referee for helping to improve the clarity of the paper.
WCGH appreciates the use of the computer facilities at the Kavli
Institute for Particle Astrophysics and Cosmology.
WCGH is supported by NASA LTSA grant NAG5-13032.
The work of AYP is supported in part by FASI (Rosnauka) grant
NSh-2600.2008.2 and by RFBR grants 05-02-22003 and 08-02-00837.
The work of GC is supported in part by CNRS grant PICS 3202.
 

\end{document}